# Gravity model for dyadic Olympic competitions

Hyeseung Choi[1], Hyungsoo Woo[1], Ji-Hyun Kim[2], Jae-Suk Yang[1*]

[1] Graduate School of Future Strategy, KAIST, Daejeon 34141, Republic of Korea
[2] School of Business, Yonsei University, Seoul 03722, Republic of Korea

**Abstract**

In the Olympic Games, professional athletes representing their nations compete regardless of economic, political and cultural differences. In this study, we apply gravity model to observe characteristics, represented by 'distances' among nations that directly compete against one another in the Summer Olympics. We use dyadic data consisting of medal winning nations in the Olympic Games from 1952 to 2016. To compare how the dynamics changed during and after the Cold War period, we partitioned our data into two time periods (1952–1988 and 1992–2016). Our research is distinguishable from previous studies in that we newly introduce application of gravity model in observing the dynamics of the Olympic Games. Our results show that for the entire study period, countries that engaged each other in competition in the finals of an Olympic event tend to be similar in economic size. After the Cold War, country pairs that compete more frequently tend to be similar in genetic origin.

# 1. Introduction

Athletes who represent diverse nations compete fiercely in the Olympic Games. The significance of winning a medal at the Olympic Games goes beyond individual achievement; it is also a triumph for the countries that the athletes represent [1]. The Olympic Games, in which many nations around the world participate regardless of differing political ideologies, cultural practices and economic status, is one of the few international competitions held consistently despite the geopolitical circumstances of the time. It affords us a unique opportunity to examine dynamics between competing nations during and after the Cold War.

The results of the Olympics are analysed in diverse disciplines. In studies of economic resources and the Olympic Games, the focus is mainly on relations between economic resources and medal totals [2]–[8]. In another literature in measuring the host effect of the Olympics, which is coined as the 'Olympic effect', uses gravity model to analyse the host country advantage in terms of exports [9].

In this study, we apply the gravity model to unveil the embedded patterns that competing nations have in the Olympic Games. Using the Olympic Games data, we herein focus on the relations between nations that actually competed against one another in the Olympic Games. From the gravity model, we adopt the notion of the mass and distances. As much as the total population is one of the most important factors that determines Olympic participation [2, 3] and that it is commonly used to represent the mass in literatures [10, 11], we use total population of any two competing nations as the mass. We address the question of who competes against whom in the Olympic Games, observing similarities between countries that directly compete against one another in the Olympic Games. In our case, we observe similarities in terms of GDP per capita and genetic origin which are introduced as distances. Our approach to this matter is to focus on pairwise competition between nations that actually competed at the individual event level in every Summer Olympic Games since 1952.

In section 2, we present operationalization of our data using bipartite network projection. Then we explain our network-based approach as well as theoretical background and its applications of the gravity model, which derived from Newton's law of gravitation. In section 3, we move towards our analysis, which consists of two main parts. First, we take network-based approach by utilizing maximum spanning trees (MST) algorithm to create a general topological overview of the Olympic network. Then, we conduct a modularity Q test to observe community cohesion by continent, in terms of polity, GDP and population. These two procedures justify the use of the gravity model-based analysis. In the second part of our analysis, we conduct a PPML regression analysis using the gravity model with data from 1952 to the recent 2016 Rio Olympic Games. Last section includes discussions of our findings and implications.

# 2. Methodology and data

*2.1. Bipartite projection*

For each Olympic game held since 1952, we obtained a list of medalist nations for each event-gender of the Olympic Games during our study period. In classifying each event of the Games, there are several hierarchical categories that each event belongs to; sport (e.g. aquatics) as the largest category and discipline (e.g. swimming), event (e.g. 200m freestyle) and event-gender (e.g. 200m freestyle - male) as the descending categories. For each event-gender of each Olympic game, we made bipartite projection. The yellow and blue nodes in Fig. 1A represents event at the gender level and medalists countries respectively for each Olympic game. Links are made between two sets of nodes of those who won a medal, regardless of the medal colors. The focus in our study lies with the mere fact that countries competed against one another. Therefore, we did not take into consideration of medal colours. Next, as illustrated in Fig 1B, we projected our bipartite network of Fig. 1A into one-mode network by countries, where two country nodes are connected if they share at least one sport event node.

We excluded self-loops in our one-mode network because the focus of our analysis is to observe the relationships between different nations, not within nations. Links in our network represents the frequencies of competition between two nations. The thickness of the links accounts for weights, representing summed encounter frequencies of two nations for each Olympic Game. We did not take into consideration differences in the format of each event. Some sports, such as judo and table tennis, involve single elimination competition, while basketball and softball follow the "three-up, three-count" format. We assumed that medalist nations directly competed against one another if they received medals from the same event.

We excluded following data from our dataset. First, we excluded data from the 1980 and 1984 Olympic Games due to the bias coming from these two Games. It was clear that the major boycott by the U.S. and its 65 allies of the 1980 Moscow Olympics and the retaliatory boycott by the Eastern bloc of the 1984 Los Angeles Olympics during the Cold War period would cause biased results. Second, we excluded Olympic teams that do not represent a specific nation, such as the IOA (Independent/Individual Olympic Athletes) and EUN (Unified Team) because the level of analysis in this research is limited to nations. Likewise, we excluded unified teams where two nations competed as a single team during the Cold War period, such as the United Team of Germany (EUA).

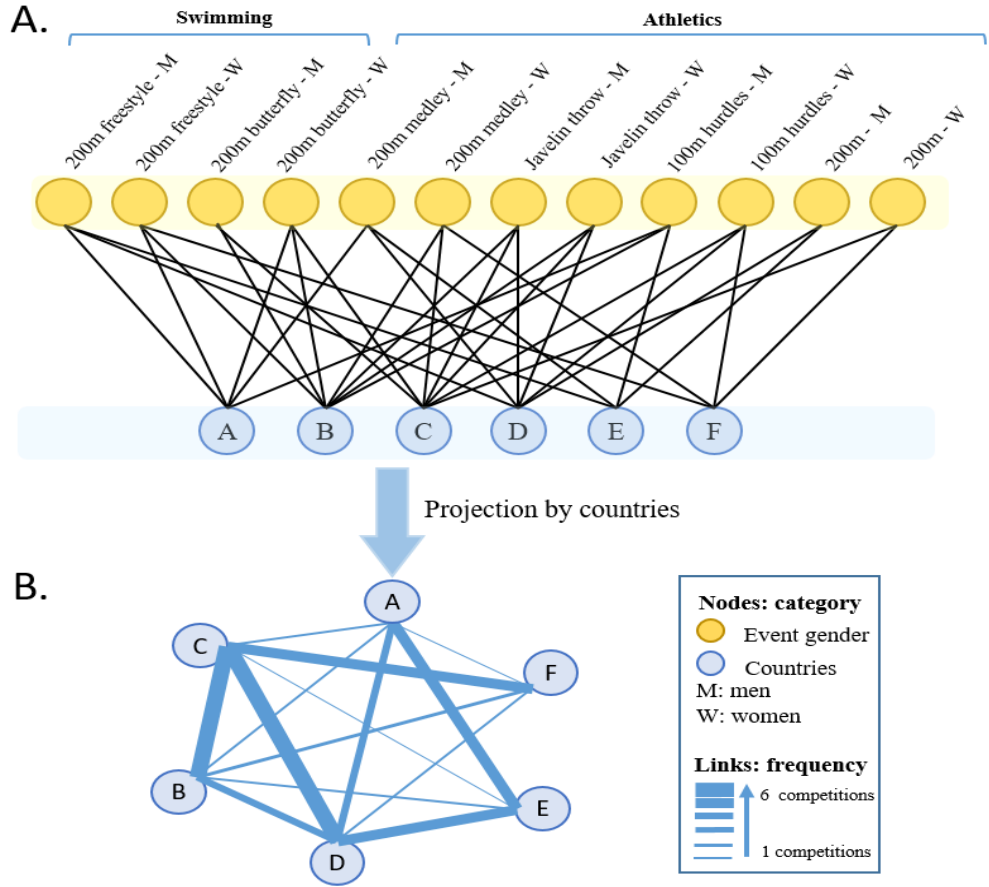

**Figure 1. Example of bipartite network projection of Olympic data**. (A) Bipartite network of two sets of nodes, event-gender (yellow) and countries (blue). (B) Projected network by country nodes. The thickness of links represents competition frequency. In this example network, country B and C engaged in 6 competitions.

*2.2. Network Topology*

       To show topological overviews of pairwise competition between nations, we use the Minimum Spanning Tree (MST) algorithm [12]. Although the MST method is useful for visualizing the basic structure of a network topology [13], some important links may be eliminated in the process, which may serve as a critical link in interpreting the network topology. Therefore, to avoid the potential problem that comes from MST algorithm, we also use modularity Q with the original network to detect cluster cohesion. Modularity Q is used to measure the structure of the network to observe how the network can be divided into modules, or clusters. If the modules are well divided into clusters, then the network is considered to have dense connections among the nodes within the cluster, while it has low connections between different clusters[14]. For both MST and modularity, we added node attributes such as

population, economy size, polity level and geographical continents of countries.

Our sample for network-based analysis consists of nations and their competing relations, represented as nodes and links, respectively. Table 1 shows descriptive statistics for Olympics network. The number of nodes and links generally increases over time observed. In 1952, there are 43 countries (nodes) that won a medal in the Olympics, while there are 86 countries in 2016 Rio Olympics. Although the number of nodes increases by two-folds, and links by approximately 2.7 folds, average degree remains similar. Given that there is an average of three medals per event, it is understandable that the density of the nodes remains similar.

**Table 1. Data composition of Olympic network**

| Year | Host city | Number of nodes | Number of links | Average degree |
|---|---|---|---|---|
| 1952 | Helsinki | 43 | 184 | 8.56 |
| 1956 | Melbourne | 37 | 139 | 7.51 |
| 1960 | Rome | 43 | 137 | 6.37 |
| 1964 | Tokyo | 40 | 145 | 7.25 |
| 1968 | Mexico City | 44 | 163 | 7.41 |
| 1972 | Munich | 48 | 168 | 7.00 |
| 1976 | Montreal | 41 | 192 | 9.37 |
| 1988 | Seoul | 49 | 229 | 9.35 |
| 1992 | Barcelona | 62 | 288 | 9.30 |
| 1996 | Atlanta | 79 | 406 | 8.84 |
| 2000 | Sydney | 80 | 444 | 9.27 |
| 2004 | Athens | 74 | 444 | 9.27 |
| 2008 | Beijing | 85 | 453 | 8.46 |
| 2012 | London | 85 | 462 | 9.27 |
| 2016 | Rio de Janeiro | 86 | 509 | 9.10 |

*2.3. Gravity model*

The gravity model originated from Newton's law of universal gravitation. The law articulates that all forms of matter in the universe are attracted to each other through the force of gravity. It is represented as follows:

$$F = G \frac{m_1 m_2}{r^2} \quad (1)$$

The force between two objects, represented by F, is proportional to the product of the two masses, $m_1$ and $m_2$, and inversely proportional to the square of the distance, *r*, between the two masses. This law of physics has also been applied to interpret social phenomena. For example, economists have applied the gravity model to measure bilateral trade flows based on the relative economic sizes and distance between nations [14-17].

The most widely used gravity model in international economics takes the following form[18]:

$$T_{ij} = \beta_0 Y_i^{\alpha_1} Y_j^{\alpha_2} D_{ij}^{\alpha_3} \eta_{ij}, \qquad (2)$$

where T denotes trade flows between countries *i* and *j*; Y represents the economic masses of countries *i* and *j*; D denotes distances between countries *i* and *j*; $\alpha_0, \alpha_1, \alpha_2, \alpha_3$ are coefficients of each parameter; and $\eta_{ij}$ is an error term. The gravity model for trade predicates that trade flows between countries *i* and *j* are proportional to the product of the sizes of the economies of the two countries, normally measured by their GDP, and inversely proportional to the distance between them. Distance is not limited to geographical distance between the two countries; it may potentially encompass all factors that may cause trade friction between two nations[19]. To measure the parameters of interest by least squares estimation, we follow the long tradition in international economics research of log-linearizing equation (2) to derive equation (3), as follows:

$$\ln T_{ij} = \ln \beta_0 + \beta_1 \ln Y_i + \beta_2 \ln Y_j - \beta_3 \ln D_{ij} + \ln \eta_{ij} \qquad (3)$$

Use of equation (3) in the gravity model framework is not limited to international economics research; it has also been widely used in different disciplines to study phenomena such as migration [19, 20], tourism [21, 22], patenting behaviour [23, 24], international arms transfers [26], and transportation [12, 26] to explain certain behaviours or characteristics affecting their respective flows. We use the applied gravity model to observe similarities among countries that engage in direct competition at the Olympic Games. There are links that can be defined by political or social phenomena, rather than geographical distance [27]; thus, we extend the notion of distance to encompass non-geographical distances in our model, such as economic and genetic distances. Also, our analysis includes other distance variables to control for factors that may affect the Olympic Games, such as linguistics and geographical proximity. To solve the problems arising from the presence of heteroscedasticity under the non-linear gravity model, we use Poisson pseudo-maximum likelihood (PPML), which provides consistent estimates of the original non-linear model [18]. Use of the Poisson regression method results in a log-linear model; it also aids in avoiding underprediction of the estimates [19].

The gravity model of dyadic Olympic competition is specified as follows:

$$W_{ijt} = ln\alpha_0 + \beta_1 \ln(POP_{it} \times POP_{jt}) + \beta_2 ln\text{GDPpc}_{ijt} + \beta_3 ln\text{GEO}_{ij} + \beta_4 ln\text{GENE}_{ij} + \beta_5 ln\text{LANG}_{ij} + \delta_1 ln\text{POLITY}_{ijt} + \delta_2 \text{COLONY}_{ijt} + \delta_3 \text{ALLIANCE}_{ijt} + \delta_4 \text{HOST}_{ijt} + \varepsilon_{ijt} \qquad (4)$$

where the dependent variable, W for weight, represents the frequency of competition between countries *i* and *j* during year *t*. Since the nature of Olympic competition does not allow differentiation between home and destination countries, it is impossible to estimate the coefficients of the origin and destination variables separately. As a solution, we used the product of the populations of any two competing nations. Thus, POP represents the product of the total populations of countries *i* and *j*, and GDPpc accounts for GDP per capita distances between countries *i* and *j* at year *t*. GEO accounts for the geographical distance between countries *i* and *j*. The remaining variables, GENE and LANG, measure different distances between paired nations. Our model also includes four dummy variables, POLITY, COLONY,

ALLIANCE and HOST. To measure the political distance between two competing nations, we use a dummy variable with a value of 1 if two nations have the same political ideology (i.e., democracy or autocracy) and 0 if otherwise. COLONY is coded as 1 if the competing nations were or have been in a colonial relationship since 1945. ALLIANCE is coded as 1 if the paired nations have a formal alliance at time $t$; HOST is coded as 1 if one of the competing nations hosts the Olympic Games at time $t$. All independent variables are logged, except for the dummy variables. A year fixed effect is included in the regression and observations are clustered at the country–pair level. Detailed description of each variable and its source is listed in Table 2.

**Table 2. Variable descriptions and sources**

| Variable | Description | Source |
| --- | --- | --- |
| Matched frequency (weight) | Competition frequency between two countries | Official IOC Website, IOC Olympic Studies Centre |
| Logged population (ln_pop) | Logged value of product of two nations' total population | Maddison Project (1952–1988) [28], World Bank Database (1992–2015) |
| Logged geographical distance (ln_geo) | Logged value of simple distance between capitals (km) | CEPII geodistance data Database [29] |
| Logged GDP per capita (ln_gdppc) | Logged value of GDP per capita | Maddison Project (1952–1988), World Bank Database (1992–2015) |
| Polity dummy (polity) | 1 if pairs share the same type of government system | Polity IV Database [30] |
| Colonial relationship dummy (colony) | 1 if pairs were ever in a colonial relationship | CEPII geodistance data [29] |
| Alliance dummy (alliance) | 1 if pairs are in formal military or strategic alliance | Correlates of War Project's Formal Alliance (v4.1) [31] |
| Logged genetic (ln_gene) | Logged value of weighted FST genetic distance | "Ancestry and Development: New Evidence." [32] |
| Logged linguistic distance (ln_lang) | Logged value of weighted linguistic distance index | "Ancestry, Language and Culture." [33] |
| Host dummy | 1 if one of the pairs hosted the Olympic Games in that year | Official IOC website |

## 3. Results and discussions

*3.1. Community clusters*

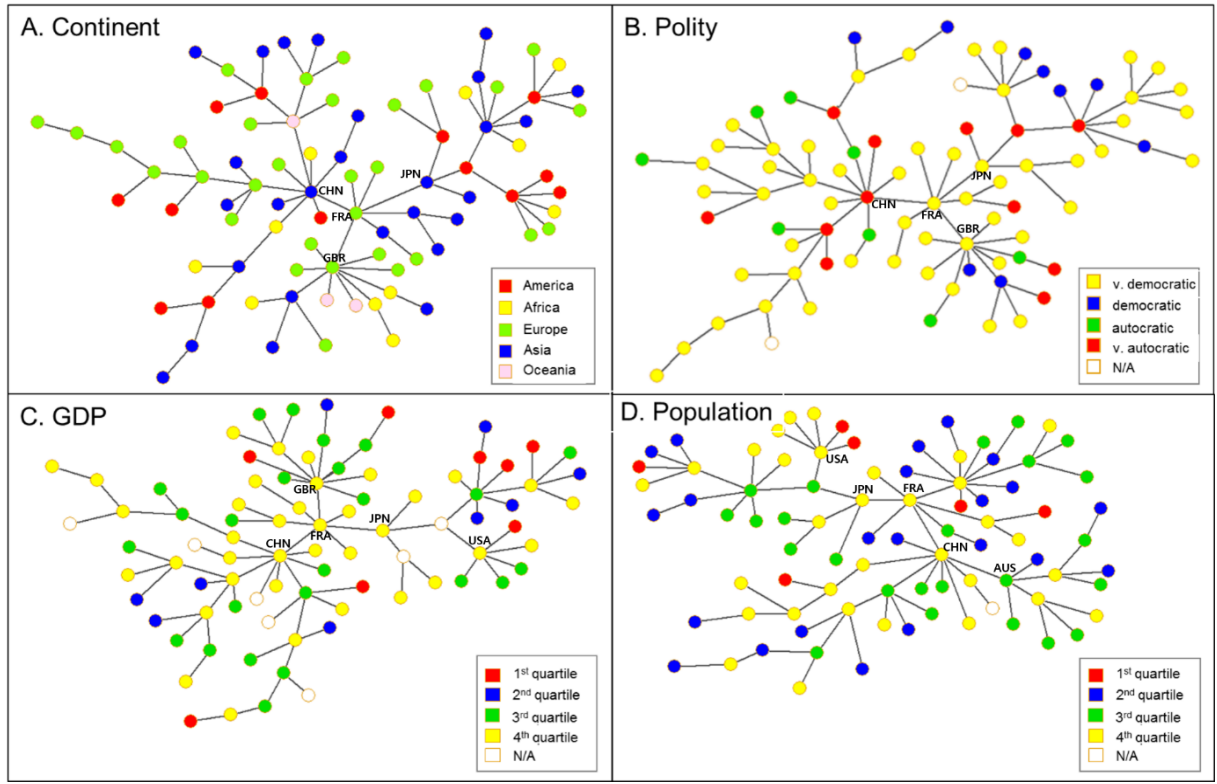

**Figure 2.** An MST of 2016 Olympics medal results composed of 86 nodes and 509 links. Node attributes are classified as stated in the legend. Each node represents a nation, and competition between two nations is represented through links. (A) Nodes are coloured according to the continent to which the country belongs ($Q$ = 0.065). (B) Node attributes represent polity, or types of government within a given country. Countries are classified into four categories ($Q$ = 0.035). (C) The GDP of each country is classified into four quartiles ($Q$ = -0.032). (D) The population is also classified into four quartiles ($Q$ = -0.037).

Fig. 2 presents an example of the MST of the 2016 Olympic Games. We expect to see clustered communities in the network, tightly connected links and only a few links formed between communities. In Fig. 2, it is difficult to distinguish clustered communities clearly by node colour. In Fig. 2A, five different continents, America, Africa, Europe, Asia and Oceania, are represented by different colours; Fig. 2B represents democracy levels of the nations; and Figs. 2C and 2D are coloured in accordance with GDP and population quartiles. In Fig. 2B, yellow nodes, which represent very democratic nations, are scattered throughout the network, but not clustered.

As can be seen in Fig. 2, modularity calculations for data in 2016 show that not all values exceed the 0.1 level. For other years, the values of modularity Q are similar to those presented in Fig. 2. Table 3 shows the results of modularity Q calculations of the post-Cold War period. We calculated our modularity Q based on four node attributes as shown in Table 3. Given that a modularity value of zero means that all nodes belong to the same community in terms of four node attributes, this finding indicates weak community cohesion.

**Table 3. Results of modularity Q calculations for the post-Cold War period**

|           | 1992  | 1996  | 2000 | 2004 | 2008  | 2012  | 2016  |
|-----------|-------|-------|------|------|-------|-------|-------|
| Continent | 0.08  | 0.06  | 0.06 | 0.06 | 0.06  | 0.04  | 0.06  |
| Democracy | -0.01 | 0.01  | 0.00 | 0.02 | 0.00  | -0.01 | 0.03  |
| GDP       | -0.01 | 0.03  | 0.02 | 0.03 | 0.03  | 0.02  | -0.03 |
| Population| -0.02 | -0.02 | 0.00 | 0.00 | -0.02 | -0.01 | -0.04 |

Weak community cohesion may result from the fact that Olympic competition outcomes represent an aggregate of many factors that influence the Games. However, modularity, by definition, only captures community cohesion for one attribute at a time. Thus, we employ the gravity model to identify the concurrent factors that influence the results of competition between paired nations in the Olympic Games.

*3.2. PPML Estimation*

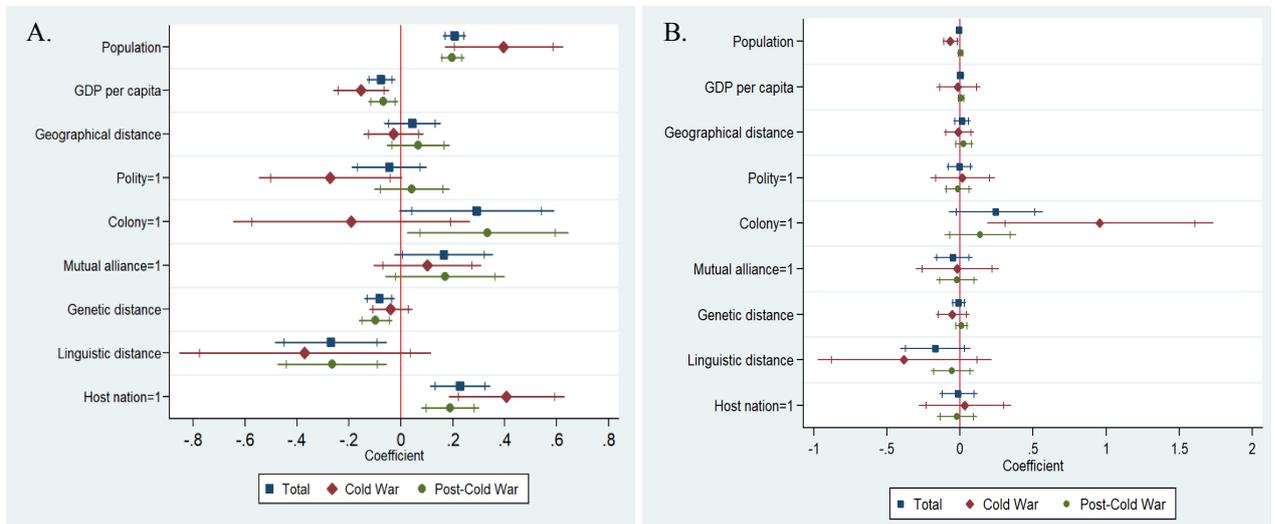

**Figure 3.** Results of the PPML estimation of pairwise Olympic competition (A) and the result of the shuffled data (B). Blue square plots represent the results for the entire study period (1952-2016), red diamond plots represent the results for the Olympic Games held during the Cold War era (1952-1988), and green circle plots represent the Olympics held during the post-Cold War period (1992-2016). Capped spikes mark confidence intervals of 90% and other spikes represent 95% confidence intervals.

Fig. 3A presents the results of the PPML estimation with corresponding coefficients ($\beta$) of equation 4 in section 2. Due to direct and indirect geopolitical influences during the Cold War period [27-30], we partitioned our dataset into two datasets for the Cold War period and post-Cold War period to facilitate comparison. Explaining the dynamics of the Olympic Games cannot be accomplished without consideration of the political factors that characterized the Games during the Cold War. For this reason, we partitioned the data into two time periods to distinguish between the Olympic Games during the Cold War and the period after the Cold War. During the study period (1952–2016), the Wald $\chi^2$ value is 191.17 (df = 23). Comparing the p-value to the alpha level (prob > $\chi^2$ = 0.000, two-tailed test), we see that for the Olympic Games held during the Cold War era (1952–1988), the Wald $\chi^2$ value is 130.67 (df = 16) with prob > $\chi^2$ = 0.000. For the post-Cold War period (1992–2016), the Wald $\chi^2$ value is 129.39 (df = 15) with prob > $\chi^2$ = 0.000.

Our study focuses on pairwise relational characteristics among competing nations. To test for potential endogeneity problem, we followed endogeneity test design that are used for dyadic data [38]. The idea behind this diagnostic is that when the dyads are randomized, then the results of the PPML estimation should lose its significance due to its randomness condition. Fig. 3B shows the results of the estimation based on shuffled data. For all three randomized results, most of the variables lose statistical significance. This indicates that the results from PPML estimations are due to the unique relationship of the Olympic competitions.

As our results show, compared to the post-Cold War period, values for most of the variables during the Cold War era are not statistically significant. This may be because during the Cold War era, the medal results at the Olympics were directly affected by major political disruptions of the time. Boycotts and disruptions occurred due to political tensions and ongoing conflicts. Thus, we can conclude that the athletes participating in the Olympic Games during the Cold War period competed under pressure from political circumstances in addition to whatever individual challenges they may have faced, which may have affected their performances and the competition patterns in the Games.

When it comes to government type, considerably significant (p-value = 0.052) differences in values for the polity variable during the Cold War period indicate that countries that competed against one another in the Olympic Games differed politically. In other words, democratic nations tended to compete more with autocratic nations, as compared to those with same political ideology. However, after the Cold War, the polity variable is no longer significant, while the value for the dummy for colonial relationships becomes significant (p-value = 0.034). This indicates that in the post-Cold War period, countries tended to compete more often if the two competing nations were in a colonial relationship.

Also noticeable is the host effect of the Olympic Games, which is significant in all time periods (p-value = 0.000 for all three estimations). This indicates that if a pair of nations includes a host country of that specific Olympic Games, the frequency of competition increased in that year. This result is also in line with those in previous studies indicating that hosting nations do receive certain advantages[3], [9], [39].

GDP per capita is statistically significant overall (p-value = 0.004) during the Cold War period (p-value = 0.005) and significant after the Cold War (p-value = 0.016). All three estimations show a negative coefficient (total: -0.076, Cold War: -0.153, post-Cold War: - 0.068). As specified in the gravity model, a negative coefficient of distance *r* indicates less distance (i.e., similarity). These results indicate that, over the entire study period, two countries that engage more in direct competition tend to be similar in GDP per capita. This not only applies to those in the top tiers of the Olympics, but also to those in the bottom tiers. In other words, even countries that only won a few medals in total in the Olympic Games tend to have competed against nations with similar economic circumstances, rather than competing against those with considerable economic differences.

While other studies acknowledge the importance of the role of GDP per capita in performance in the Olympic Games, our relational analysis showed that at the event level at the Olympics, countries competed more often against other countries with similar GDP per capita. Our analysis shows that at the event level of the Olympic Games, it is not about the size of the GDP per capita, but about how similar the competing countries are in terms of GDP per capita. Thus, our analysis adds new insight into the economic aspect of the Olympic Games.

During the Cold War period, genetic distance is not statistically significant (p-value = 0.346). On the other hand, in the post-Cold War era, genetic distance becomes very significant (p-value = 0.002) with a negative coefficient of -0.097. Interpreted as GDP per capita, this shows that in the post-Cold War period, countries tend to engage in more competition when two nations are similar in genetic origin.

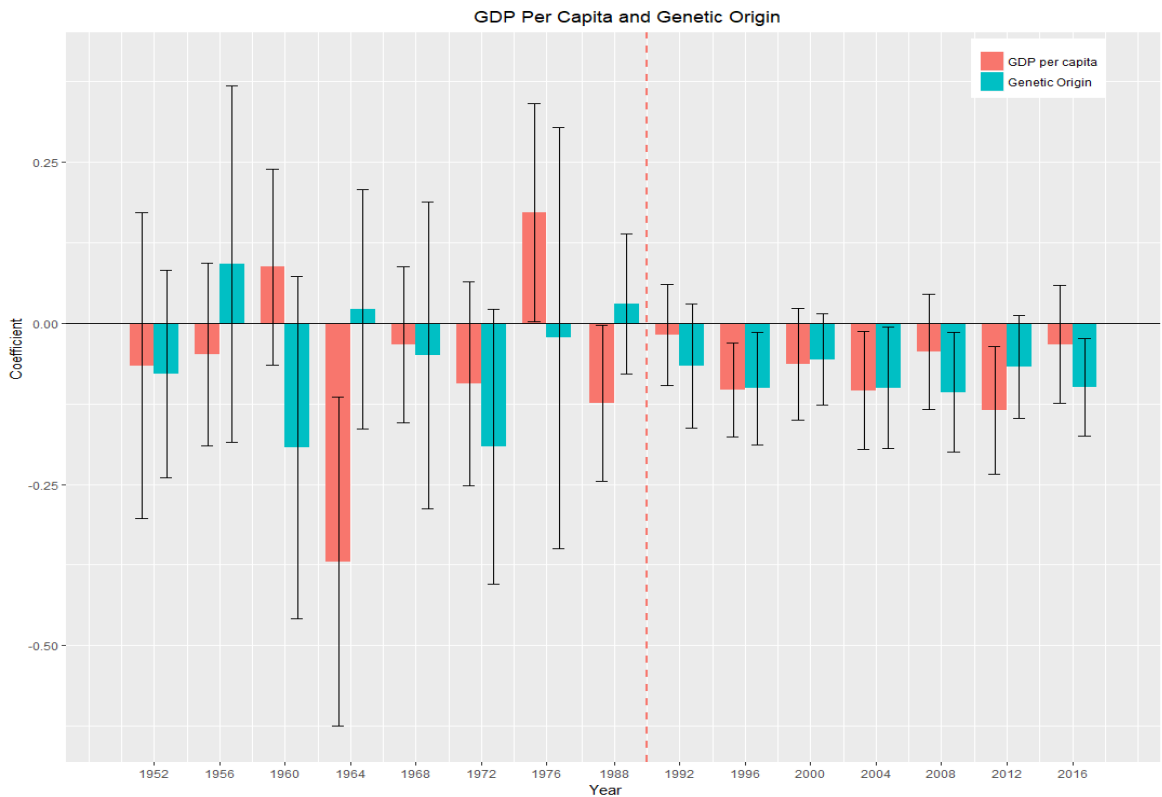

**Figure 4.** Similarities in GDP per capita and genetic origin over observed study time. Negative coefficients indicate similarities among competing nations. Vertical red dotted line marks the end of the Cold War era. During the Cold War era, the coefficient signs fluctuate for both GDP per capita and genetic origin with large confidence intervals. Comparatively, after the Cold War, coefficients become negative with less fluctuation in confidence intervals. (Error bars= 95% Confidence intervals)

In Fig. 4, we graph the results for our two main distances of interest over the 1952–2016 period. Fig. 4 shows changes in the coefficient over time, with focus on our variables of interest, GDP per capita and genetic origin. During the Cold War, coefficients of GDP per capita and genetic origin are high in magnitude, while after the Cold War, values for genetic origin and GDP per capita are negative and low in magnitude. These results reveal that there is a tendency to compete against those who are similar in these two respects.

## 4. Conclusion

We applied the gravity model to find similarities between those who engaged in direct competition in the Summer Olympic Games since 1952. This study is the first to analyse the pairwise relational similarities of competing nations at the individual event level in the Olympic Games over time. Our focus shifts away from that of conventional research where the focus is on total medal counts. With our different approach, we revealed that those directly engaging in competition tended to be similar in GDP per capita and genetic origin.

Our analysis leads to the interesting conclusion that countries who competed more often at the event level tended to be similar in GDP per capita and genetic origin after the Cold War. However, during the Cold War period, countries competed against one another when they are similar in economic size.

Our novel approach towards the Olympic competition by applying the gravity model can serve as a starting point where the focus of the study is on relational differences or similarities among competing nations. Also, we gained new insights to by introducing the notion of masses and distances of the gravity model.

There are limitations to our research due to our holistic approach towards the Olympic Games during the time period spanning from 1952 to 2016. In operationalizing our data, we only included competing pairs of nations with athletes who won medals. The results of our analysis could be improved if the data were extended to include countries with athletes who did not win a medal, examining, for example, all the pairwise competing relationships after a certain round (e.g., the quarterfinals in a single-elimination tournament). Also, our dataset does not take into consideration differences between sports. In future research, it would be interesting to observe how distances differ at the sports discipline level. Despite the limitations stemming from our method of data operationalization, this study is the first attempt of its kind to analyse the characteristics of actual competing pairs within the context of the Olympic Games.